\documentclass[letterpaper, 12pt]{article}
\usepackage{jheppub}

\usepackage{graphicx}
\usepackage{bbm}
\usepackage{amsmath}
\usepackage{amssymb}
\usepackage{amsthm}
\usepackage{physics}
\usepackage{mathrsfs}
\usepackage{afterpage}
\usepackage{rotating}
\usepackage{epstopdf}
\usepackage{float}
\usepackage{upgreek}
\usepackage{mathtools}
\usepackage{multirow}
\usepackage[labelformat=simple]{subcaption}
\usepackage{calligra}
\usepackage[T1]{fontenc}  
  
\begin{document}

\begin{titlepage}

\setcounter{page}{1} \baselineskip=15.5pt \thispagestyle{empty}
{\flushright {ITP-CAS-24-226}\\}
		
\bigskip\
		
\vspace{0.6cm}
\begin{center}
{\LARGE \bfseries On the Temperature Effects in QCD Axion\vspace{0.24cm}\\ Mass Mixing}
\end{center}
\vspace{0.15cm}
			
\begin{center}
{\fontsize{14}{30}\selectfont Hai-Jun Li}
\end{center}
\begin{center}
\vspace{0.25 cm}
\textsl{Key Laboratory of Theoretical Physics, Institute of Theoretical Physics, Chinese Academy of Sciences, Beijing 100190, China}\\

\vspace{-0.1 cm}				
\begin{center}
{E-mail: \textcolor{blue}{\tt {lihaijun@itp.ac.cn}}}
\end{center}	
\end{center}
\vspace{0.6cm}
\noindent

In this work, we extend the QCD axion mass mixing in the early Universe and investigate the temperature effects in the mixing.
We explore the scenario where two $Z_{\mathcal N}$ QCD axions undergo mass mixing during the QCD phase transition, yielding three distinct mixing scenarios: the mixing~I, II, and III. 
These scenarios are realized through fine-tuning of the axion decay constants, the temperature parameters, as well as the value of $\mathcal N$.
We conduct a thorough analysis of the level crossing phenomena in these three mixing scenarios, detailing the conditions under which they occur. 
Notably, in the mixing~I and II, the level crossing precedes the critical temperature of the QCD phase transition ($T_{\rm QCD}$), with minimal non-essential discrepancies in the cosmological evolution of the mass eigenvalues at $T_{\rm QCD}$. 
In contrast, the mixing~III exhibits a unique double level crossings, occurring both before and at $T_{\rm QCD}$.
Despite superficial similarities in axion evolution between the mixing~II and III, we uncover fundamental differences between them. 
Additionally, we briefly address the transition in energy density between the two axions within our mixing scenarios. 
This work contributes to a deeper understanding of the role of the QCD axion in the early Universe and its potential implications for cold dark matter.
			
\vspace{2cm}
			
\bigskip
\noindent\today
\end{titlepage}
			
\setcounter{tocdepth}{2}
			
\hrule
\tableofcontents			
\bigskip\medskip
\hrule
\bigskip\bigskip
 

\section{Introduction}

The QCD axion holds the remarkable potential to address two fundamental issues simultaneously: the strong CP problem and the mystery of dark matter (DM).
It was originally proposed by the Peccei-Quinn (PQ) mechanism \cite{Peccei:1977hh, Peccei:1977ur} with a spontaneously broken global $\rm U(1)_{PQ}$ symmetry to solve the strong CP problem in the Standard Model (SM) \cite{Weinberg:1977ma, Wilczek:1977pj, Kim:1979if, Shifman:1979if, Dine:1981rt, Zhitnitsky:1980tq, Berezhiani:1989fp}.
It acquires a minuscule mass from QCD non-perturbative effects \cite{tHooft:1976rip, tHooft:1976snw}.
When the QCD instanton generates the axion potential, the axion can stabilize at the CP conservation minimum, which dynamically solves the strong CP problem.
See ref.~\cite{Hook:2018dlk} for a review.

On the other hand, the QCD axion is an attractive candidate for the cold DM that can be non-thermally produced via the misalignment mechanism \cite{Preskill:1982cy, Abbott:1982af, Dine:1982ah}.
The oscillation of coherent axion state in the potential in the early Universe can contribute to the overall DM.
The equation of motion (EOM) for this state is that of a damped harmonic oscillator, in which the friction term is proportional to the Hubble parameter $H(T)$ and the spring constant is proportional to the axion mass $m_a(T)$. 
As the temperature decreases, the QCD axion gains a non-zero mass during the QCD phase transition and begins to oscillate when $m_a(T)$ is comparable to $H(T)$, the system thus changes from an overdamped to an underdamped oscillator and the oscillation explains the observed cold DM abundance.
See also refs.~\cite{Marsh:2015xka, DiLuzio:2020wdo, OHare:2024nmr} for recent reviews.
By assuming the $\sim\mathcal{O}(1)$ initial misalignment angle, it gives the upper limit on the classical QCD axion window $f_a\lesssim 10^{11}-10^{12}\, {\rm GeV}$, where $f_a$ is the axion decay constant.
While the lower limit $f_a\gtrsim 10^9\, {\rm GeV}$ is given by SN\,1987A \cite{Raffelt:1987yt, Turner:1987by, Mayle:1987as} and neutron star cooling \cite{Leinson:2014ioa, Hamaguchi:2018oqw, Buschmann:2021juv}.

In recent years, there have been several attempts to broaden the above classical QCD axion window, one of which is the lighter-than-usual axion model, the $Z_{\mathcal N}$ QCD axion \cite{Hook:2018jle, DiLuzio:2021pxd}.
In this scenario, the ${\mathcal N}$ mirror worlds that are nonlinearly realized by the axion field under a discrete $Z_{\mathcal N}$ symmetry can coexist in Nature.
The zero-temperature $Z_{\mathcal N}$ axion mass is exponentially suppressed at the QCD phase transition critical temperature due to the suppressed non-perturbative effects on the axion potential.
The $Z_{\mathcal N}$ axion can also simultaneously solve the strong CP problem and account for the DM.
In order to solve the strong CP problem, here $\mathcal N$ should be an odd number and greater than or equal to 3.  
In addition, this lighter QCD axion can account for the cold DM through the trapped+kinetic misalignment mechanism \cite{DiLuzio:2021gos, Co:2019jts}.  

The Universe with a large number of these axions, the QCD axions or axion-like particles (ALPs), is called the axiverse \cite{Arvanitaki:2009fg, Cicoli:2012sz, Demirtas:2021gsq}.
In the axiverse, it is natural to take into account the cosmological evolution of the multiple axions.
In recent years the mass mixing in the multiple axions model \cite{Hill:1988bu} has attracted extensive attention.
If considering the non-zero mass mixing between two axions, the nontrivial cosmological evolution process of the mass eigenvalues called the level crossing can take place during the QCD phase transition, such as the mixing between the QCD axion and ALP \cite{Daido:2015cba}, between the QCD axion and sterile axion \cite{Cyncynates:2023esj}, and between the $Z_{\mathcal N}$ axion and ALP \cite{Li:2023uvt}, etc.
In general, the level crossing can occur before the QCD phase transition critical temperature $T_{\rm QCD}$.
But this is not always the case, as pointed out in ref.~\cite{Li:2023uvt}, there may still be a second level crossing at $T_{\rm QCD}$ in the mixing with a $Z_{\mathcal N}$ axion, which is called the double level crossings.
The level crossing can lead to the adiabatic transition of the axion energy density, which is similar to the MSW effect \cite{Wolfenstein:1977ue, Mikheyev:1985zog, Mikheev:1986wj} in neutrino oscillations\footnote{It should be noted that the latter also involves neutrino flavor swapping after the crossing.}.
It also has some intriguing cosmological implications, including the modification of axion relic density and isocurvature perturbations \cite{Kitajima:2014xla, Ho:2018qur, Cyncynates:2021xzw, Li:2023xkn}, the formation of domain walls \cite{Daido:2015bva}, the effects on gravitational waves and primordial black holes \cite{Cyncynates:2022wlq, Chen:2021wcf, Li:2024psa}, and the composition of dark energy \cite{Muursepp:2024mbb, Muursepp:2024kcg}, etc.

So far, the aforementioned axion mass mixing can be summarized as a mixing between an axion possessing a temperature-dependent mass and one with a constant mass, but what emerges if considering two axions in a theory, both of which exhibit the temperature-dependent masses?
In ref.~\cite{Li:2024okl}, they recently investigated this mixing case with two QCD axions, one canonical QCD axion and one $Z_{\mathcal N}$ axion, showing a novel level crossing phenomenon in the context of mass mixing.

In this work, we extend this QCD axion mass mixing and investigate the temperature effects in the mixing.
Our motivation lies in exploring a novel mixing mechanism that can enhance our understanding of axion behavior in the early Universe and potentially offer a new explanation for the generation of axion DM.
We consider the mass mixing between two $Z_{\mathcal N}$ axions during the QCD phase transition, resulting in three mixing scenarios, the mixing~I, II, and III.
This can be accomplished by fine-tuning the model parameters.
The level crossing in these three mixing scenarios and the conditions for them to occur are investigated in detail.
In the mixing~I and II, the level crossing can take place before $T_{\rm QCD}$, while the cosmological evolution of the mass eigenvalues at $T_{\rm QCD}$ has a minor non-essential difference.
In the mixing~III, we find that the double level crossings can take place before and at $T_{\rm QCD}$, respectively.
Although the axion evolution in the mixing~II and III appears similar, they have fundamental differences.
Finally, we briefly discuss the axion energy density transition in the mixing. 
Our mixing scenarios may also have some intriguing cosmological implications.
 
The rest of this paper is structured as follows. 
In section~\ref{sec_review_on_QCD_axion}, we briefly review the QCD axion.
In section~\ref{sec_Extended_QCD_axion_mass_mixing}, we investigate the extended QCD axion mass mixing, focusing on three mixing scenarios.
Finally, the conclusion is given in section~\ref{sec_Conclusion}.
Additionally, a brief description of the derivation of the axion mixing potential is included in appendix~\ref{appendix_1}.
 
\section{Brief review on QCD axion}
\label{sec_review_on_QCD_axion}
 
In this section, we briefly review the QCD axion.
We first introduce the canonical QCD axion.
The effective potential of the canonical QCD axion, arising from QCD non-perturbative effects, is given by \cite{Marsh:2015xka}
\begin{eqnarray}
V_{\rm QCD}(\phi)=m_a^2(T) f_a^2\left[1-\cos\left(\dfrac{\phi}{f_a}\right)\right]\, ,
\end{eqnarray}
where $\phi$ is the axion field, $f_a$ is the axion decay constant, and $\theta=\phi/f_a$ is the axion angle.
The $m_a(T)$ is the temperature-dependent QCD axion mass \cite{Marsh:2015xka}
\begin{eqnarray}
m_a(T)=
\begin{cases}
\dfrac{m_\pi f_\pi}{f_a}\dfrac{\sqrt{z}}{1+z}\, , & T\leq T_{\rm QCD}\\ 
\dfrac{m_\pi f_\pi}{f_a}\dfrac{\sqrt{z}}{1+z}\left(\dfrac{T}{T_{\rm QCD}}\right)^{-b}\, , & T> T_{\rm QCD} 
\end{cases} 
\label{mQCDT}
\end{eqnarray} 
where $T_{\rm QCD}\simeq150\, \rm MeV$ is the critical temperature of the QCD phase transition, $m_\pi$ and $f_\pi$ are the mass and decay constant of the pion, respectively, $z\equiv m_u/m_d\simeq0.48$ is the ratio of the up to down quark masses, and $b\simeq4.08$ is an index taken from the dilute instanton gas approximation \cite{Borsanyi:2016ksw}.
The first term in eq.~(\ref{mQCDT}) corresponds to the zero-temperature QCD axion mass $m_{a,0}^{\rm QCD}$.
The QCD axion can account for the DM through the misalignment mechanism.
As the cosmic temperature decreases, the axion starts to oscillate when its mass $m_a(T)$ becomes comparable to the Hubble parameter $H(T)$.
If considering the pre-inflationary scenario in which the PQ symmetry is spontaneously broken during inflation, the current QCD axion DM abundance is given by \cite{Li:2023xkn}
\begin{eqnarray}
\Omega_ah^2\simeq0.14 \left(\dfrac{g_{*s}(T_0)}{3.94}\right)\left(\dfrac{g_*(T_{\rm osc})}{61.75}\right)^{-0.42}\left(\dfrac{f_a}{10^{12}\, \rm GeV}\right)^{1.16}\left\langle\theta_i^2f(\theta_i)\right\rangle\, ,
\end{eqnarray}
where $h\simeq0.68$ is the reduced Hubble constant, $T_0$ is the current cosmic microwave background (CMB) temperature, $T_{\rm osc}$ is the axion oscillation temperature, and $f(\theta_i)$ is the anharmonicity factor \cite{Lyth:1991ub, Visinelli:2009zm}.
In order to explain the observed cold DM abundance, $\Omega_{\rm DM}h^2\simeq0.12$ \cite{Planck:2018vyg}, we have the $\sim\mathcal{O}(1)$ initial misalignment angle \cite{Li:2023xkn}
\begin{eqnarray}
\theta_i\simeq0.87\left(\dfrac{g_{*s}(T_0)}{3.94}\right)^{-1/2}\left(\dfrac{g_*(T_{\rm osc})}{61.75}\right)^{0.21}\left(\dfrac{f_a}{10^{12}\, \rm GeV}\right)^{-0.58}\, .
\end{eqnarray}
Notice that if the PQ phase transition occurs during inflation, the spikes in the spectrum of density fluctuations may result in an excessive production of primordial black holes. 
Moreover, if the PQ phase transition occurs after inflation, one must consider the influence of axion string networks as well as large-scale correlations within the energy density of axion field oscillations \cite{Khlopov:1999tm}. 
 
\begin{figure}[t]
\centering
\includegraphics[width=0.70\textwidth]{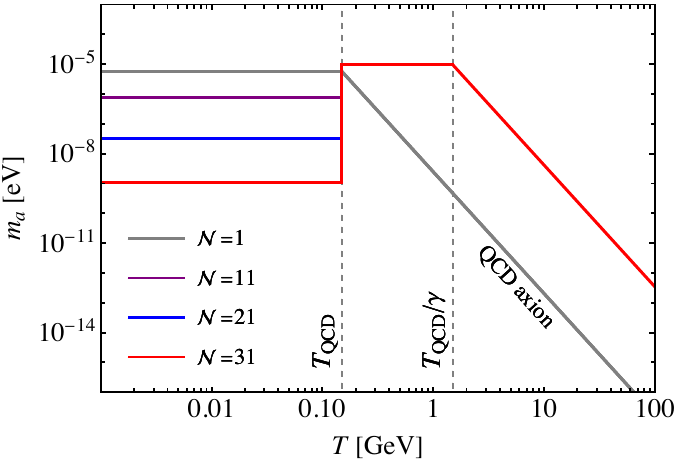}
\caption{The temperature-dependent $Z_{\mathcal N}$ axion mass $m_a(T)$ as a function of the cosmic temperature $T$.
The different colored lines represent the different values of $\mathcal N$, and they merge into the red line at $T\geq T_{\rm QCD}$.
The vertical gray lines represent the temperatures $T_{\rm QCD}$ and $T_{\rm QCD}/\gamma$, respectively.
Notice also that the case $\mathcal N=1$ (gray solid line) corresponds to the canonical QCD axion [eq.~(\ref{mQCDT})].
Here we set $f_a=10^{12}\, \rm GeV$ and $\gamma=0.1$.}
\label{fig_znmaT}
\end{figure}  

In the $Z_{\mathcal N}$ axion scenario \cite{Hook:2018jle, DiLuzio:2021pxd, DiLuzio:2021gos}, the $\mathcal N$ copies of the SM worlds that are nonlinearly realized by the axion field under a $Z_{\mathcal N}$ symmetry ($0\leqslant k \leqslant {\mathcal N}-1$)
\begin{eqnarray}
{\rm SM}_k&\to&{\rm SM}_{k+1}\, ,  \\
\phi&\to&\phi+\dfrac{2\pi k}{\mathcal N}f_a\, ,
\end{eqnarray}
can coexist with the same coupling strengths as that in the SM \cite{DiLuzio:2021pxd}
\begin{eqnarray}
\mathcal L=\sum_{k=0}^{\mathcal N-1}\left[\mathcal L_{{\rm SM}_k} +\dfrac{\alpha_s}{8\pi}\left(\dfrac{\phi}{f_a}+\dfrac{2\pi k}{\mathcal N}\right)G_k\widetilde{G}_k\right]+\cdots\, ,
\end{eqnarray}
where $\mathcal L_{{\rm SM}_k}$ corresponds to the copies of the SM Lagrangian, $G_k \widetilde{G}_k$ is the topological term, and $\alpha_s$ is the strong fine structure constant. 
The $Z_{\mathcal N}$ axion has a nontrivial temperature-dependent potential.
In the large $\mathcal N$ limit, the total axion potential can be given in all generality by \cite{DiLuzio:2021pxd}
\begin{eqnarray}
V_{\mathcal N}(\phi)\simeq -\dfrac{m_a^2(T) f_a^2}{{\mathcal N}^2} \cos\left({\mathcal N} \dfrac{\phi}{f_a}\right)\, ,
\end{eqnarray}
and the temperature-dependent $Z_{\mathcal N}$ axion mass can be described by \cite{DiLuzio:2021gos}
\begin{eqnarray}
m_a(T)\simeq
\begin{cases}
\dfrac{m_\pi f_\pi}{\sqrt[4]{\pi} f_a}\sqrt[4]{\dfrac{1-z}{1+z}}{\mathcal N}^{3/4}z^{{\mathcal N}/2}\, , &T\leq T_{\rm QCD}\\
\dfrac{m_\pi f_\pi}{f_a}\sqrt{\dfrac{z}{1-z^2}}\, , &T_{\rm QCD} < T \leq \dfrac{T_{\rm QCD}}{\gamma}\\
\dfrac{m_\pi f_\pi}{f_a}\sqrt{\dfrac{z}{1-z^2}}\left(\dfrac{\gamma T}{T_{\rm QCD}}\right)^{-b}\, , &T>\dfrac{T_{\rm QCD}}{\gamma}
\end{cases} 
\label{massZNT}
\end{eqnarray}
where $\gamma\in(0,1)$ is a temperature parameter\footnote{Note that this temperature parameter defines the high-temperature and intermediate-temperature regions. Specifically, it can be expressed as $\gamma_{\rm min} < \gamma < \gamma_{\rm max}$, corresponding respectively to the mirror copies of the SM that are the coolest and the hottest \cite{DiLuzio:2021gos}. For simplicity, we denote it as $\gamma$ here.}.
The first and second terms in eq.~(\ref{massZNT}) can also be defined as the masses $m_{a,0}$ and $m_{a,\pi}$, respectively.

In figure~\ref{fig_znmaT}, we show the $Z_{\mathcal N}$ axion mass $m_a(T)$ as a function of the temperature $T$.
The different colored lines correspond to the different values of $\mathcal N$, merging into the red line as the temperature exceeds $T_{\rm QCD}$.
Compared with the canonical QCD axion [eq.~(\ref{mQCDT})], we have a relation between their zero-temperature axion masses 
\begin{eqnarray}
\dfrac{m_{a,0}}{m_{a,0}^{\rm QCD}}= \dfrac{1}{\sqrt[4]{\pi}}\sqrt[4]{\dfrac{\left(1-z\right)\left(1+z\right)^3}{z^2}}{\mathcal N}^{3/4}z^{\mathcal N/2}\, ,
\end{eqnarray}
which is suddenly exponentially suppressed at $T_{\rm QCD}$ due to the $Z_{\mathcal N}$ symmetry.
The $Z_{\mathcal N}$ axion can also account for the DM through the trapped+kinetic misalignment mechanism, see refs.~\cite{DiLuzio:2021gos, Co:2019jts} for more details.
Notice also that the effects of $Z_{\mathcal N}$ axion on the stellar configuration of white dwarfs were recently investigated in ref.~\cite{Balkin:2022qer}.
They found that the modifications of the mass-radius relationship of white dwarfs provide the opportunity to explore vast, uncharted regions of parameter space, without the necessity for the $Z_{\mathcal N}$ axion to constitute cold DM.

\section{Extended QCD axion mass mixing}
\label{sec_Extended_QCD_axion_mass_mixing}

In this section, we extend the QCD axion mass mixing and investigate the temperature effects in the mixing.
 
\subsection{The model}

\begin{figure}[t]
\centering
\includegraphics[width=0.70\textwidth]{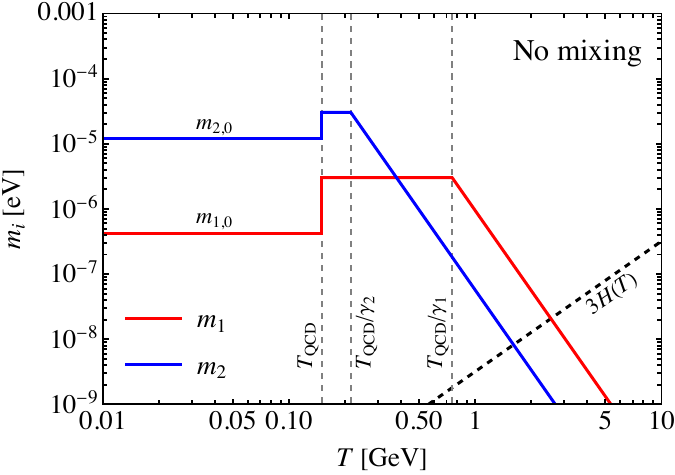}
\caption{The illustration of the no mixing case.
The red and blue solid lines represent the temperature-dependent axion masses $m_1(T)$ and $m_2(T)$, respectively.
The three vertical gray lines represent the temperatures $T_{\rm QCD}$, $T_{\rm QCD}/\gamma_2$, and $T_{\rm QCD}/\gamma_1$, respectively.
The black dashed line represents the Hubble parameter $H(T)$.
Notice that we have assumed $\gamma_1<\gamma_2$.
The model parameters can be found in table~\ref{tab_1}.}
\label{fig_miT}
\end{figure}

For our purpose, we consider a multiple QCD axions model, two $Z_{\mathcal N}$ axions $\phi_i$ ($i=1, 2$), with the low-energy effective Lagrangian
\begin{eqnarray}
\begin{aligned}
\mathcal{L}&\supset\dfrac{1}{2}\sum_{i=1, 2}\partial_\mu\phi_i\partial^\mu\phi_i-\dfrac{m_1^2(T) f_1^2}{{\mathcal N}_1^2} \left[1-\cos\left({\mathcal N}_1\dfrac{\phi_1}{f_1}\right)\right]\\
&-\dfrac{m_2^2(T) f_2^2}{{\mathcal N}_2^2}\left[1-\cos\left({\mathcal N}_1\dfrac{\phi_1}{f_1}+{\mathcal N}_2\dfrac{\phi_2}{f_2}\right)\right]\, ,
\label{Lagrangian_Extended}
\end{aligned}
\end{eqnarray}
where $m_i(T)$ are the temperature-dependent axion masses corresponding to $\phi_i$, and $f_i$ are the axion decay constants.
In appendix~\ref{appendix_1}, we provide a brief description of the derivation of the axion mixing potential in our model.
This framework may be realized within the Type IIB string axiverse \cite{Cicoli:2012sz, Conlon:2006tq}, encompassing two sets of ${\mathcal N}\gg 1$ copies of the SM, each possessing a $Z_{\mathcal N}$ symmetry.
The axion mass $m_i(T)$ can be described by
\begin{eqnarray}
m_i(T)=
\begin{cases}
\dfrac{m_\pi f_\pi}{\sqrt[4]{\pi} f_i}\sqrt[4]{\dfrac{1-z}{1+z}}{\mathcal N}_i^{3/4}z^{{\mathcal N}_i/2}\, , &T\leq T_{\rm QCD}\\
\dfrac{m_\pi f_\pi}{f_i}\sqrt{\dfrac{z}{1-z^2}}\, , &T_{\rm QCD} < T \leq \dfrac{T_{\rm QCD}}{\gamma_i}\\
\dfrac{m_\pi f_\pi}{f_i}\sqrt{\dfrac{z}{1-z^2}}\left(\dfrac{\gamma_i T}{T_{\rm QCD}}\right)^{-b}\, , &T>\dfrac{T_{\rm QCD}}{\gamma_i}
\end{cases} 
\label{massiT}
\end{eqnarray}
where $\gamma_i\in(0,1)$ are the temperature parameters, and we focus on a scenario that\footnote{Notice that one can also assume $\gamma_1>\gamma_2$, which will not affect our main results.}  
\begin{eqnarray}
\dfrac{T_{\rm QCD}}{\gamma_1}>\dfrac{T_{\rm QCD}}{\gamma_2} \Longrightarrow \gamma_1<\gamma_2\, .
\end{eqnarray} 
The first term in eq.~(\ref{massiT}) corresponds to the zero-temperature axion mass
\begin{eqnarray}
m_{i,0}=\dfrac{m_\pi f_\pi}{\sqrt[4]{\pi} f_i}\sqrt[4]{\dfrac{1-z}{1+z}}{\mathcal N}_i^{3/4}z^{{\mathcal N}_i/2}\, ,
\end{eqnarray}
which is exponentially suppressed due to the $Z_{\mathcal N}$ symmetry. 
The second term in eq.~(\ref{massiT}) corresponds to the mass
\begin{eqnarray}
m_{i,\pi}=\dfrac{m_\pi f_\pi}{f_i}\sqrt{\dfrac{z}{1-z^2}}\, .
\end{eqnarray}
To begin with, it is essential to illustrate the distribution of axion masses $m_i$ in the scenario where no mixing occurs, as depicted in figure~\ref{fig_miT}. 
For this purpose, we have selected two representative distributions for $m_1$ and $m_2$, which are illustrated by the red and blue lines, respectively.
See table~\ref{tab_1} for the model parameters.

Then plugging the whole mixing potential of eq.~(\ref{Lagrangian_Extended}) into the EOM
\begin{eqnarray}
\ddot\phi_i+3H\dot\phi_i+\dfrac{\partial V_{\rm mix}(\phi_i)}{\partial \phi_i}=0\, ,
\end{eqnarray}
we derive the EOM of $\phi_i$ 
\begin{eqnarray}
\ddot{\phi_2}+3H\dot{\phi_2}+\dfrac{m_2^2(T) f_2}{{\mathcal N}_2}\sin\left({\mathcal N}_1\dfrac{\phi_1}{f_1}+{\mathcal N}_2\dfrac{\phi_2}{f_2}\right)=0 \, , 
\end{eqnarray}
\begin{eqnarray}
\ddot{\phi_1}+3H\dot{\phi_1}+ \dfrac{m_1^2(T) f_1}{{\mathcal N}_1}\sin\left({\mathcal N}_1\dfrac{\phi_1}{f_1}\right)+\dfrac{{\mathcal N}_1 m_2^2(T) f_2^2}{{\mathcal N}_2^2 f_1} \sin\left({\mathcal N}_1\dfrac{\phi_1}{f_1}+{\mathcal N}_2\dfrac{\phi_2}{f_2}\right)=0\, , 
\end{eqnarray}
where one dot represents a derivative with respect to the physical time $t$, two dots represent two derivatives with respect to the time, $\partial_{\phi_i} V_{\rm mix}$ represents the derivative of $V_{\rm mix}$ with respect to $\phi_i$, and $H(T)$ is the Hubble parameter.
If considering the oscillation amplitudes of $\phi_i$ are much smaller than the corresponding decay constants $f_i$, we can obtain the mass mixing matrix in our model
\begin{eqnarray}
\mathbf{M}^2=
\left(
\begin{array}{cc}
m_2^2(T)  & \quad \dfrac{{\mathcal N}_1 m_2^2(T) f_2}{{\mathcal N}_2 f_1} \\
\dfrac{{\mathcal N}_1 m_2^2(T) f_2}{{\mathcal N}_2 f_1} & \quad m_1^2(T)+\dfrac{{\mathcal N}_1^2 m_2^2(T) f_2^2}{{\mathcal N}_2^2 f_1^2}
\end{array}
\right)\, .
\end{eqnarray}
Diagonalizing the mass mixing matrix, we derive the heavy and light mass eigenstates $a_h$ and $a_l$, respectively,
\begin{eqnarray}
\left(\begin{array}{c}
a_l \\
a_h 
\end{array}\right)
=\left(\begin{array}{cc}
\cos \alpha & \quad \sin \alpha \\
-\sin \alpha  & \quad   \cos \alpha
\end{array}\right)
\left(\begin{array}{c}
\phi_2 \\
\phi_1 
\end{array}\right)\, ,
\end{eqnarray} 
with the mass mixing angle $\alpha$
\begin{eqnarray}
\cos^2\alpha(T)=\dfrac{1}{2}\left(1+\dfrac{m_1^2(T)-m_2^2(T)+\dfrac{{\mathcal N}_1^2 m_2^2(T) f_2^2}{{\mathcal N}_2^2 f_1^2}}{m_h^2(T)-m_l^2(T)}\right) \, ,
\end{eqnarray} 
where $m_h(T)$ and $m_l(T)$ (assuming $m_h>m_l$) are the corresponding mass eigenvalues
\begin{eqnarray}
\begin{aligned}
m_{h,l}^2(T)&=\dfrac{1}{2}\left[m_1^2(T)+m_2^2(T)+\dfrac{{\mathcal N}_1^2 m_2^2(T) f_2^2}{{\mathcal N}_2^2 f_1^2}\right]\\
&\pm\dfrac{1}{2{\mathcal N}_2^2 f_1^2}\bigg[-4 {\mathcal N}_2^4 m_1^2(T) m_2^2(T) f_1^4\\
&+\bigg({\mathcal N}_1^2 m_2^2(T) f_2^2 + {\mathcal N}_2^2 f_1^2\left(m_1^2(T)+m_2^2(T)\right)\bigg)^2\bigg]^{1/2}\, .
\end{aligned}
\end{eqnarray}  

\begin{table}[t]
\centering
\begin{tabular}{lccccccc}
\hline\hline
Scenarios   &    $f_1$ ($\rm GeV$)   &    $\gamma_1$   &  ${\mathcal N}_1$  &   $f_2$ ($\rm GeV$)  &  $\gamma_2$  &  ${\mathcal N}_2$   &  $T_\times$ ($\rm MeV$) \\
\hline
No mixing  &   $10^{12.5}$     &  0.2   & 9 &   $10^{11.5}$  &  0.7 &  5  & --- \\
Mixing~I &    $10^{12.5}$     &  0.2   & 9 &   $10^{11.5}$  &  0.7 &  5    & 381.28 \\
Mixing~II &   $10^{12.5}$     &  0.2   & 9 &   $10^{11.5}$  &  0.7 &  13  & 377.45 \\
Mixing~III &   $10^{12.5}$     &  0.2   & 9 &   $10^{11.5}$  &  0.7 &  21 & 377.04 \\
 \hline\hline
\end{tabular}
\caption{The typical model parameters for $f_i$, $\gamma_i$, and ${\mathcal N}_i$ used in this work.
Notice that the axion decay constants $f_1$ and $f_2$ are chosen based on the benchmark value of $10^{12}\, {\rm GeV}$ within the classical QCD axion window $10^{9}-10^{12}\, {\rm GeV}$. 
The temperature parameters $\gamma_1$ and $\gamma_2$ are selected within the range of 0 to 1. 
The values of ${\mathcal N}_1$ and ${\mathcal N}_2$ are arbitrarily chosen positive odd numbers, provided that they meet the conditions relevant to different mixing scenarios.
In order to make clear comparisons, we attempt to minimize alterations to the model parameters while keeping other parameters unchanged.
Therefore, in the simplest case, we only need to fine-tune ${\mathcal N}_2$ to achieve our illustrative purpose.
Additionally, we present the values of $T_\times$ in the mixing scenarios.}
\label{tab_1}
\end{table}

\subsection{Level crossing in the mass mixing}

In this subsection, we investigate the level crossing phenomenon in the context of the aforementioned extended QCD axion mass mixing.
Since we have previously set $\gamma_1<\gamma_2$, the subsequent results and discussions are predicated on this assumption. 

\begin{figure}[t]
\centering
\includegraphics[width=0.70\textwidth]{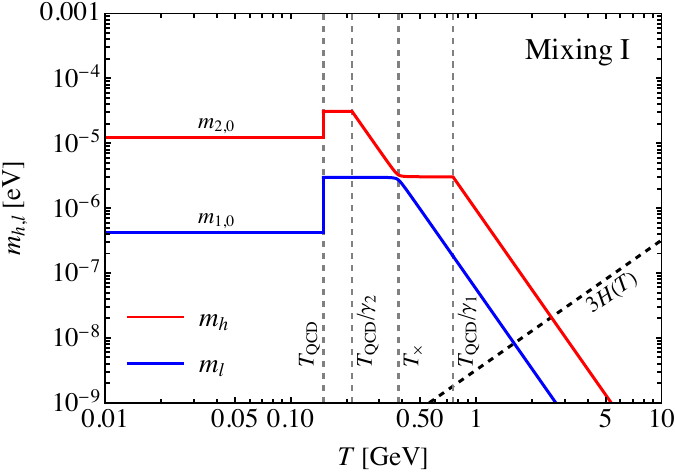}
\caption{The illustration of the mixing~I.
The red and blue solid lines represent the temperature-dependent mass eigenvalues $m_h(T)$ and $m_l(T)$, respectively.
The third vertical gray line represents the level crossing temperature $T_\times$.
The model parameters can be found in table~\ref{tab_1}.}
\label{fig_mhlT_1}
\end{figure}

\subsubsection{The mixing~I}

We first consider a mixing scenario that the zero-temperature axion mass $m_{2,0}$ of $\phi_2$ is larger than the mass $m_{1,\pi}$ of $\phi_1$,
\begin{eqnarray}
\dfrac{m_\pi f_\pi}{\sqrt[4]{\pi} f_2}\sqrt[4]{\dfrac{1-z}{1+z}}{\mathcal N}_2^{3/4}z^{{\mathcal N}_2/2} > \dfrac{m_\pi f_\pi}{f_1}\sqrt{\dfrac{z}{1-z^2}}\, ,
\end{eqnarray}
then we have
\begin{eqnarray}
\dfrac{f_2}{f_1}<\dfrac{1}{\sqrt[4]{\pi}}\sqrt[4]{\dfrac{\left(1+z\right)\left(1-z\right)^3}{z^2}}{\mathcal N}_2^{3/4}z^{{\mathcal N}_2/2} \simeq 0.73\times0.48^{{\mathcal N}_2/2} {\mathcal N}_2^{3/4}\, .
\label{con_1}
\end{eqnarray}
See figure~\ref{fig_mhlT_1} for the detailed illustration of this particular mixing scenario.
The red and blue lines correspond to the temperature-dependent mass eigenvalues $m_h(T)$ and $m_l(T)$, respectively, as they vary with respect to the cosmic temperature $T$.
For the purpose of comparison with a scenario where no mixing occurs, here we use the same set of model parameters as those employed in figure~\ref{fig_miT}.

In this case, the level crossing can take place when the difference of $m_h^2(T)-m_l^2(T)$ gets a minimum value.
By solving 
\begin{eqnarray}
\dfrac{d\left(m_h^2(T)-m_l^2(T)\right)}{dT}=0\, ,
\end{eqnarray}
we have the level crossing temperature in the mixing
\begin{eqnarray}
T_\times= \dfrac{T_{\rm QCD}}{\gamma_2} \left(\dfrac{\left({\mathcal N}_1^2 f_2^2 + {\mathcal N}_2^2 f_1^2\right)^2}{{\mathcal N}_2^4 f_1^2 f_2^2 - {\mathcal N}_1^2 {\mathcal N}_2^2 f_2^4}\right)^{\frac{1}{2b}} \, , 
\label{Tx}  
\end{eqnarray} 
which will last for a parametric duration \cite{Li:2023xkn}
\begin{eqnarray}
\Delta T_\times=\bigg|\dfrac{1}{\cos\alpha(T)}\dfrac{d\cos\alpha(T)}{d T}\bigg|^{-1}_{T=T_\times}\, ,
\label{Delta_Tx}
\end{eqnarray}  
corresponding to the time $\Delta t_\times$.
Notice also that from eq.~(\ref{Tx}) we can obtain the following relation
\begin{eqnarray}
{\mathcal N}_2^4 f_1^2 f_2^2 - {\mathcal N}_1^2 {\mathcal N}_2^2 f_2^4 > 0 \Longrightarrow \dfrac{f_2}{f_1}<\dfrac{{\mathcal N}_2}{{\mathcal N}_1}\, .
\label{con_0}
\end{eqnarray}
Given our underlying assumption that $\gamma_1<\gamma_2$, it is important to highlight that the level crossing temperature $T_\times$ remains unaffected by $\gamma_1$.
See also figure~\ref{fig_mhlT_1}, where the third vertical gray line indicates $T_\times\simeq 381.28 \, \rm MeV$. 
Our focus now shifts to exploring the temperature-dependent behavior of axions within this mixing framework. 
Initially, at high temperatures, the heavy mass eigenstate $a_h$ is composed of the axion $\phi_1$, whereas the light mass eigenstate $a_l$ is composed of $\phi_2$. 
As the cosmic temperature decreases, a fascinating phenomenon emerges: the mass eigenvalues $m_h(T)$ and $m_l(T)$ converge towards each other at $T_\times$ and then subsequently diverge. 
This convergence and subsequent divergence are collectively termed as the level crossing.
After this level crossing, a reversal in the composition of $a_h$ and $a_l$ occurs, with $a_h$ now comprising $\phi_2$ and $a_l$ consisting of $\phi_1$. 
This level crossing can lead to a significant adiabatic transition in the energy density distribution between $\phi_1$ and $\phi_2$ at $T_\times$. 

\begin{figure}[t]
\centering
\includegraphics[width=0.70\textwidth]{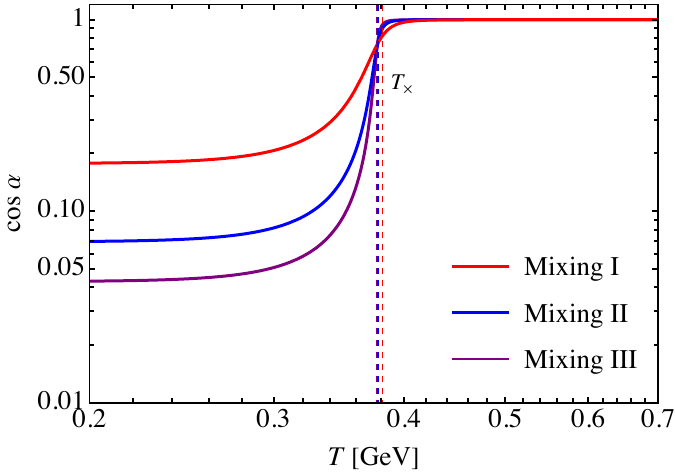}
\caption{The mass mixing angle $\cos\alpha$ as a function of the cosmic temperature $T$.
The red, blue, and purple solid lines correspond to the mixing~I, II, and III, respectively.
The vertical lines represent the corresponding level crossing temperatures $T_\times$.
The model parameters can be found in table~\ref{tab_1}.}
\label{fig_angle123}
\end{figure}

In addition to this, we show the mass mixing angle $\cos\alpha$ in figure~\ref{fig_angle123} with the red solid line.
At high temperatures, the mixing angle remains invariant, with $\cos\alpha=1$ ($\alpha=0$), indicating a stable mixing state. 
However, upon reaching the level crossing temperature $T_\times$, $\cos\alpha$ undergoes a rapid decline and eventually stabilizes at a relatively small value. 
This change in $\cos\alpha$ further emphasizes the dynamic nature of the axion mixing process and its relationship with temperature.

\subsubsection{The mixing~II}

\begin{figure}[t]
\centering
\includegraphics[width=0.70\textwidth]{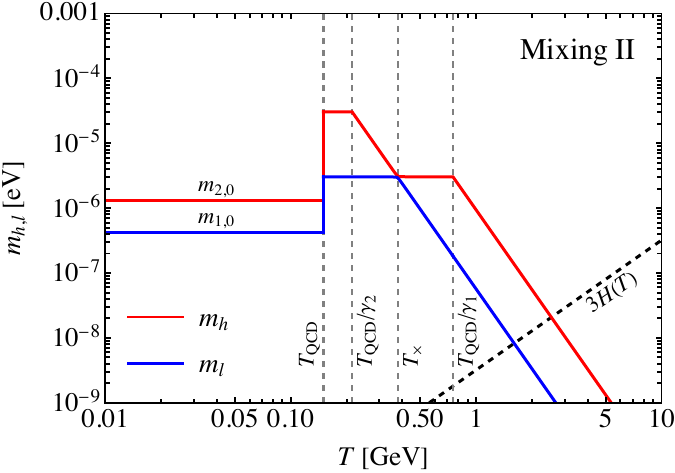}
\caption{Same as figure~\ref{fig_mhlT_1} but for the mixing~II.}
\label{fig_mhlT_2}
\end{figure}

When considering the scenario of mixing~I, it becomes apparent that it is natural to envision a situation in which ${\mathcal N}_2$ attains a sufficiently large value. 
This condition ensures that the zero-temperature axion mass $m_{2,0}$ is smaller than $m_{1,\pi}$.
Consequently, this prompts us to investigate another scenario. 
In this scenario, we consider a case where the zero-temperature axion mass $m_{2,0}$ of $\phi_2$ is smaller than the mass $m_{1,\pi}$ of $\phi_1$, but larger than the zero-temperature axion mass $m_{1,0}$ of $\phi_1$,
\begin{eqnarray}
\dfrac{m_\pi f_\pi}{\sqrt[4]{\pi} f_1}\sqrt[4]{\dfrac{1-z}{1+z}}{\mathcal N}_1^{3/4}z^{{\mathcal N}_1/2} < \dfrac{m_\pi f_\pi}{\sqrt[4]{\pi} f_2}\sqrt[4]{\dfrac{1-z}{1+z}}{\mathcal N}_2^{3/4}z^{{\mathcal N}_2/2} < \dfrac{m_\pi f_\pi}{f_1}\sqrt{\dfrac{z}{1-z^2}}\, ,
\end{eqnarray}
then we have
\begin{eqnarray}
0.73\times0.48^{{\mathcal N}_2/2} {\mathcal N}_2^{3/4} < \dfrac{f_2}{f_1} < 0.48^{({\mathcal N}_2-{\mathcal N}_1)/2}\left(\dfrac{{\mathcal N}_2}{{\mathcal N}_1}\right)^{3/4}\, .
\label{con_21}
\end{eqnarray}
Notice that there is an additional relation that should be satisfied first\footnote{This relation can be naturally satisfied in the mixing~I, since in that scenario the zero-temperature axion mass $m_{2,0}$ of $\phi_2$ is inherently larger than the mass $m_{1,\pi}$ of $\phi_1$.}, the mass $m_{1,\pi}$ of $\phi_1$ should be smaller than the mass $m_{2,\pi}$ of $\phi_2$,
\begin{eqnarray}
\dfrac{m_\pi f_\pi}{f_1}\sqrt{\dfrac{z}{1-z^2}} < \dfrac{m_\pi f_\pi}{f_2}\sqrt{\dfrac{z}{1-z^2}} \Longrightarrow \dfrac{f_2}{f_1} <1\, .
\label{con_22}
\end{eqnarray}
See also figure~\ref{fig_mhlT_2} for the illustration of this mixing scenario.
The model parameters can be found in table~\ref{tab_1}.
The axion evolution in this scenario closely resembles that observed in the mixing~I. 
However, it is worth noting that a subtle, albeit non-critical, difference emerges at the temperature $T_{\rm QCD}$. 
It is important to emphasize that the axion evolution at $T_{\rm QCD}$ does not conform to the double level crossings scenario as proposed in ref.~\cite{Li:2023uvt}. 
Instead, in the current scenario, the level crossing occurs solely at $T_\times$.
Furthermore, we show in figure~\ref{fig_angle123} the evolution of the mass mixing angle for this case. 
When compared with the mixing~I, it is evident that the mixing angle undergoes a more significant change and ultimately stabilizes at a smaller value of $\cos\alpha$. 
 
In fact, there exists non-essential difference\footnote{Here, when we refer to ``non-essential,$"$ we mean that both the mixing~I and II involve only one level crossing. In other words, there is no second level crossing occurring at $T_{\rm QCD}$.} between the scenarios mixing~I and II, and their respective axion evolution processes are similar. 
However, despite this similarity, for the purpose of facilitating a comparison with another scenario that will be discussed later, it is deemed necessary to present this scenario as a separate entity here. 
By doing so, we aim to provide a clearer and more comprehensive understanding of the various axion mixing scenarios and their implications.

\subsubsection{The mixing~III}

\begin{figure}[t]
\centering
\includegraphics[width=0.70\textwidth]{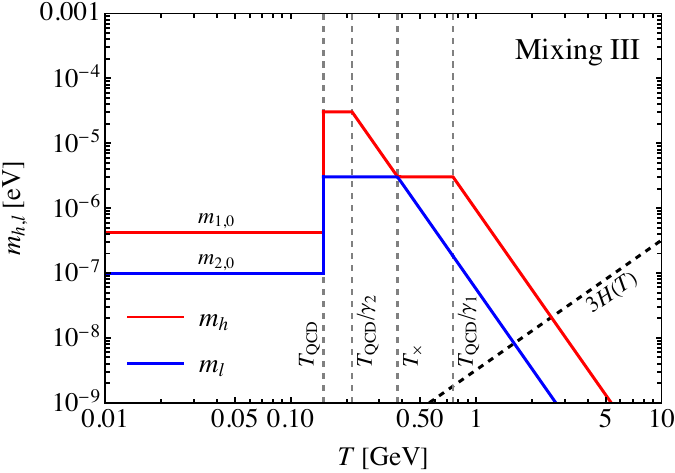}
\caption{Same as figure~\ref{fig_mhlT_1} but for the mixing~III.
Notice that the first and second level crossings can take place at $T_\times$ and $T_{\rm QCD}$, respectively.}
\label{fig_mhlT_3}
\end{figure}

As we previously discussed, if the value of ${\mathcal N}_2$ continues to increase, then we will observe the emergence of a novel level crossing phenomenon\footnote{It is worth emphasizing that, despite being described as ``novel,$"$ the fundamental nature of this level crossing phenomenon remains consistent with that of the double level crossings \cite{Li:2023uvt}.}.
Consequently, here we further consider the third mixing scenario that the zero-temperature axion mass $m_{2,0}$ of $\phi_2$ is smaller than the zero-temperature axion mass $m_{1,0}$ of $\phi_1$,
\begin{eqnarray}
\dfrac{f_2}{f_1} > 0.48^{({\mathcal N}_2-{\mathcal N}_1)/2}\left(\dfrac{{\mathcal N}_2}{{\mathcal N}_1}\right)^{3/4}\, .
\label{con_3}
\end{eqnarray}
Notice that eq.~(\ref{con_22}) should also be satisfied here; otherwise, the phenomenon we discuss here will not occur.
We show in figure~\ref{fig_mhlT_3} the illustration of this mixing scenario.
The model parameters are also listed in table~\ref{tab_1}.

By comparing figures~\ref{fig_mhlT_2} with \ref{fig_mhlT_3}, we find that they appear similar at first glance, but in reality they possess fundamental differences.
A clarification regarding the evolution of axions in this particular scenario is necessary.
At high temperatures, the heavy mass eigenstate $a_h$ is comprised of the axion $\phi_1$, whereas the light mass eigenstate $a_l$ consists of $\phi_2$.
As the cosmic temperature decreases, the first level crossing occurs at $T_\times$.
At this point, the mass eigenvalues $m_h(T)$ and $m_l(T)$ approach each other and then diverge.
Subsequently, the $a_h$ becomes comprised of $\phi_2$ and the $a_l$ becomes comprised of $\phi_1$, remaining so until $T_{\rm QCD}$ is reached.
So far, the evolution of axions follows a trajectory similar to that observed in the aforementioned mixing scenarios mixing~I and II. 
However, a significant deviation arises at the temperature $T_{\rm QCD}$.
The second level crossing takes place at $T_{\rm QCD}$, marking a pivotal turning point.
Consequently, the heavy eigenstate $a_h$ once again becomes composed of $\phi_1$, while the light eigenstate $a_l$ resumes its composition of $\phi_2$. 
This intriguing phenomenon, characterized by two distinct level crossings (at $T_\times$ and $T_{\rm QCD}$), constitutes the main difference between the mixing~II and III.

Moreover, the transition of axion energy density in this scenario will also undergo drastic alterations.
The mass mixing angle $\cos\alpha$ in this scenario is depicted in figure~\ref{fig_angle123}, denoted by the purple line.
The significant changes in axions' composition and energy density, as discussed above, reflect the complexity and diversity of their evolution within this particular framework.

\begin{figure}[t]
\centering
\includegraphics[width=0.70\textwidth]{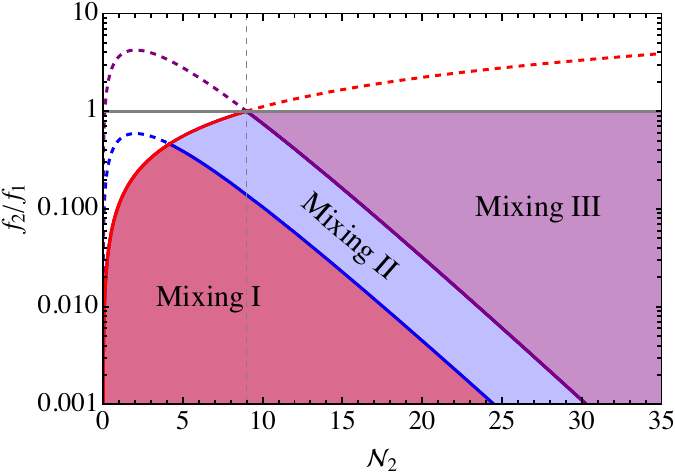}
\caption{The conditions for the mixing~I, II, and III to occur in the $\{{\mathcal N}_2,\, f_2/f_1\}$ plane.
The red, blue, and purple shaded regions correspond to the mixing~I, II, and III, respectively.
The gray solid line represents $f_2/f_1=1$. 
The vertical gray line represents ${\mathcal N}_2=9$.
Here we set ${\mathcal N}_1=9$.
Note also that ${\mathcal N_1},\, {\mathcal N_2}=2k+1$, $k\in\mathbb{N}^+$.}
\label{fig_condit123}
\end{figure}

We close this subsection by discussing the conditions for the occurrence of the above three mixing scenarios.
These conditions serve as the cornerstone for understanding the dynamics of these mixing processes.
For comparison, we plot them in the $\{{\mathcal N}_2,\, f_2/f_1\}$ plane; see figure~\ref{fig_condit123} with the shaded regions distinguishing the different mixing conditions.
The red, blue, and purple shaded regions correspond to the mixing~I, II, and III, respectively.
Furthermore, we add lines on the plot to delineate the boundaries of these shaded regions: the red line is given by eq.~(\ref{con_0}), the blue line by eq.~(\ref{con_1}), the purple line by eq.~(\ref{con_21}), and the gray solid line by eq.~(\ref{con_22}), respectively.
Notice that the vertical gray line represents a special case where ${\mathcal N}_2={\mathcal N}_1=9$.
This line serves as a reference point for understanding how changes in ${\mathcal N}_1$ affect the mixing conditions.
The parameter ${\mathcal N}_1$ plays a pivotal role in determining the conditions for all three mixing scenarios. 
To visualize this, imagine shifting the vertical gray line laterally across the plot to different values of ${\mathcal N}_1$. 
As you do so, you will observe that the shaded regions, representing the mixing conditions, shift accordingly. 
This demonstrates how variations in ${\mathcal N}_1$ can profoundly influence the mixing behavior of the scenarios in question.

\subsection{Transition of axion energy density}
 
In this subsection, we briefly discuss the energy density transition between the two axions within our mixing scenarios. 

In the context of mass mixing, the adiabatic transition can occur if the comoving axion numbers of the eigenstates $a_h$ and $a_l$ are both conserved independently at the level crossing temperature \cite{Ho:2018qur, devaud:hal-00197565}.
This conservation is crucial in determining the transition that may take place.
The condition for adiabatic transition is given by \cite{Ho:2018qur}
\begin{eqnarray}
\Delta t_\times \gg \max\left[\dfrac{2\pi}{m_l(T)}\bigg|_{T=T_\times}, \, \dfrac{2\pi}{m_h(T)-m_l(T)}\bigg|_{T=T_\times}\right]\, ,
\label{condition_adiabatic}
\end{eqnarray}
corresponding to the temperature given in eq.~(\ref{Delta_Tx}).
One can also numerically verify this inequality to determine whether the adiabatic condition is met at $T_\times$.
 
Here we only illustrate how the axion energy density is transferred in the mixing, without intending to discuss further details. 
Since the axion evolution in the mixing~I and II is similar, their energy density transitions are also comparable. 
Therefore, we will focus solely on discussing the mixing~II.
In this scenario, at high temperatures the axion fields $\phi_1$ and $\phi_2$ are frozen at arbitrary initial misalignment angles, and start to oscillate at the corresponding oscillation temperatures $T_{\rm osc}$.
Then during the period from $T_{\rm osc}$ to $T_\times$, the axion energy density is adiabatic invariant with a comoving number.
If the condition for adiabatic transition is satisfied at $T_\times$, the energy density of $\phi_1$ will be transferred to $\phi_2$ and vice versa.
Subsequently, the final axion energy density can be further calculated through the trapped+kinetic misalignment mechanism.

In the mixing~III, the axion energy transition before the QCD phase transition critical temperature $T_{\rm QCD}$ is similar to that discussed above, but it begins to diverge at $T_{\rm QCD}$.
Since the second level crossing can occur at $T_{\rm QCD}$, then the $a_h$ will comprise $\phi_1$ and the $a_l$ will comprise $\phi_2$ once again.
Consequently, the final axion energy density will differ based on the corresponding zero-temperature axion masses.
Notice that the axion energy density at the second level crossing is non-adiabatic.

Finally, we briefly comment on the impact of our mixing scenarios on axion relic density, in comparison to simply altering the initial misalignment angles.
Since we are considering two $Z_{\mathcal N}$ axions here, the final abundance of each single axion will either be suppressed or enhanced --- in general, one will be suppressed, while the other will be enhanced\footnote{This may be used to explain the overproduction problem of QCD axion DM abundance, as well as to address the insufficient QCD axion abundance in the dark dimension \cite{Li:2024jko}.} --- depending on the specific choice of our mixing potential and other model parameter settings. 
Similarly, as discussed in ref.~\cite{Li:2023uvt} regarding changes in $Z_{\mathcal N}$ axion abundance in the cases of single/double level crossings, we can naively estimate that the variation in axion abundance in this work should follow a pattern similar to $\sim f_i\theta_i$ for the same zero-temperature axion mass. 
Note that we have not considered the difference in axion mass at the first oscillation, as this difference will be minimal in this work, depending on the variation between the temperature parameters $\gamma_i$.
Therefore, we estimate that the effect of our mixing scenarios, when the difference in $\gamma_i$ is insignificant, is on a similar scale as that of simply altering the initial misalignment angles.

\section{Conclusion}
\label{sec_Conclusion}

In summary, we have extended the concept of QCD axion mass mixing in the early Universe and investigated the temperature effects within this mixing framework.
We first reviewed the canonical QCD axion and the lighter $Z_{\mathcal N}$ QCD axion.
Then we extended the investigation to include the temperature effects on the QCD axion mass mixing.
Finally, we briefly discussed the energy density transition between the two axions in the mixing. 

For our purpose, we consider the mass mixing between two $Z_{\mathcal N}$ axions during the QCD phase transition, resulting in three mixing scenarios, the mixing~I, II, and III.
This can be accomplished by fine-tuning the model parameters, such as the axion decay constants $f_i$, the temperature parameters $\gamma_i$, and also the numbers of ${\mathcal N}_i$.
The level crossing in these three mixing scenarios and the conditions for their occurrence in the $\{{\mathcal N}_2,\, f_2/f_1\}$ plane are studied and illustrated in detail.

In the scenarios mixing~I and II, the level crossing can take place before the QCD phase transition critical temperature $T_{\rm QCD}$, while the cosmological evolution of the mass eigenvalues at $T_{\rm QCD}$ exhibits a minor non-essential difference.
In order to compare with the mixing~III, we list the mixing~II separately.
In the scenario mixing~III, we find that the double level crossings can take place before and at $T_{\rm QCD}$, respectively.
Although the axion evolution in the mixing~II and III appears similar, there exist fundamental distinctions between them --- the latter involves a second level crossing, whereas the former does not.

In the mixing~II (or I), the axion evolution from high to low temperatures can be approximately\footnote{Notice that this approximation of axion evolution is valid in the case where there is a significant difference in the axion decay constants $f_i$.} described by $\phi_1\to\phi_2$ and $\phi_2\to\phi_1$, whereas in the mixing~III, the axion evolution is characterized by $\phi_1\to\phi_2\to\phi_1$ and $\phi_2\to\phi_1\to\phi_2$.
Consequently, the final axion energy density will differ based on the corresponding zero-temperature axion masses.
We find that these mixing scenarios have the potential to simultaneously resolve the issues of QCD axion DM overproduction and underproduction under different circumstances.

Beyond its potential role as cold DM, the extended framework of QCD axion mass mixing presented in our scenarios may also yield other intriguing cosmological implications, such as isocurvature fluctuations, dark energy, domain walls, gravitational waves, and primordial black holes, warranting further research and investigation.
Therefore, this study holds substantial theoretical significance and may provide new directions and insights for future experimental observations and cosmological research. 
 
\section*{Acknowledgments}

We thank Yu-Feng Zhou for helpful discussions.
This work was supported by the Key Laboratory of Theoretical Physics in Institute of Theoretical Physics, CAS.

\appendix

\section{Derivation of axion mixing potential}
\label{appendix_1}

Here we provide a brief description of the derivation of the axion mixing potential for this work, while similar discussions can also be found in the appendix of ref.~\cite{Li:2024okl}.
Considering a general two-axion ($\phi_1$ and $\phi_2$) mixing potential, we have
\begin{eqnarray}
V_{\rm mix}=\Lambda_1^4\left[1-\cos\left(n_{11}\dfrac{\phi_1}{f_1}+n_{12}\dfrac{\phi_2}{f_2}\right)\right]+\Lambda_2^4\left[1-\cos\left(n_{21}\dfrac{\phi_1}{f_1}+n_{22}\dfrac{\phi_2}{f_2}\right)\right]\, ,
\end{eqnarray}
where $\Lambda_i$ are the overall scales, and $n_{ij}$ are the domain wall numbers.
To explore the emergence of the level crossing phenomenon, we must set the domain wall numbers, ensuring that one is set to zero, while the others are not.
This consideration and procedure are justifiable, and numerous pieces of literature that explore the mixing of two axions have adopted similar methodologies.
Additionally, setting the CP phases to zero can be seen as equivalent to requiring an independent solution for addressing the strong CP problem, as this hypothesis neither affects the evolution of the two axion fields during their mixing nor changes the axion energy density in the process. 

Let's return to our work on the mixing of the two $Z_{\mathcal{N}}$ axions.
Given that both axions are charged under two copies of the SM and coupled to QCD, we expect that $\phi_1$ will receive the dominant contribution from the QCD potential. 
Consequently, this will lead to the cancellation of the potential typically observed in the case of the conventional $Z_{\mathcal N}$ axion, whereas no such cancellation will take place for $\phi_2$.
For our purposes, we need to consider the following parameter settings
\begin{eqnarray}
n_{11}={\mathcal N}_1\, ,\quad n_{12}=0\, ,\quad n_{21}={\mathcal N}_1\, ,\quad n_{22}={\mathcal N}_2\, .
\end{eqnarray}
Hence, we derive the mixing potential as presented in eq.~(\ref{Lagrangian_Extended}), formulated as follows
\begin{eqnarray}
V_{\rm mix}=\dfrac{m_1^2(T) f_1^2}{{\mathcal N}_1^2} \left[1-\cos\left({\mathcal N}_1\dfrac{\phi_1}{f_1}\right)\right]+\dfrac{m_2^2(T) f_2^2}{{\mathcal N}_2^2}\left[1-\cos\left({\mathcal N}_1\dfrac{\phi_1}{f_1}+{\mathcal N}_2\dfrac{\phi_2}{f_2}\right)\right]\, .
\end{eqnarray}
Notice that the first term aligns perfectly with the single $Z_{\mathcal N}$ axion potential, although this alignment is not an absolute requirement within this framework.
Alternatively, since the two $Z_{\mathcal N}$ axions here can be considered equivalent, the parameter settings can also be configured as follows
\begin{eqnarray}
n_{11}={\mathcal N}_1\, ,\quad n_{12}={\mathcal N}_2\, ,\quad n_{21}=0\, ,\quad n_{22}={\mathcal N}_2\, .
\end{eqnarray}
Then the selection of other model parameters (specifically, $f_i$, $\gamma_i$, and ${\mathcal N}_i$) necessitates a thorough and independent analysis; however, this does not undermine or contradict the primary findings presented in this work.

\bibliographystyle{JHEP}
\bibliography{references}

\end{document}